\documentclass[fleqn,usenatbib]{mnras}

\usepackage{newtxtext,newtxmath}
\usepackage{float}
\usepackage{ulem}
\usepackage{subfigure}
\usepackage[T1]{fontenc}

\DeclareRobustCommand{\VAN}[3]{#2}
\let\VANthebibliography\thebibliography
\def\thebibliography{\DeclareRobustCommand{\VAN}[3]{##3}\VANthebibliography}

\usepackage{graphicx}	
\usepackage{amsmath}	

\title[2D watershed void clustering]{2D watershed void clustering for probing the cosmic large-scale structure}

\author[Y. Song et al.]{
Yingxiao Song$^{1,2}$,
Yan Gong$^{1,2,3}$\thanks{Email:gongyan@bao.ac.cn},
Qi Xiong$^{1,2}$,
Kwan Chuen Chan$^{6,7}$,
Xuelei Chen$^{1,2,4,5}$,
Qi Guo$^{1,2}$,\newauthor
Yun Liu$^{1,2}$, 
and Wenxiang Pei$^{1,2}$
\\\\
$^{1}$National Astronomical Observatories, Chinese Academy of Sciences,20A Datun Road, Beijing 100012, China\\
$^{2}$School of Astronomy and Space Sciences, University of Chinese Academy of Sciences(UCAS),Yuquan Road NO.19A Beijing 100049, China\\
$^{3}$Science Center for China's Space Survey Telescope, National Astronomical Observatories, Chinese Academy of Sciences,\\20A Datun Road, Beijing 100101, China\\
$^{4}$Department of Physics, College of Sciences, Northeastern University, Shenyang 110819, China\\
$^{5}$Centre for High Energy Physics, Peking University, Beijing 100871, China\\
$^{6}$School of Physics and Astronomy, Sun Yat-sen University, 2 Daxue Road, Tangjia, Zhuhai, 519082, China\\
$^{7}$CSST Science Center for the Guangdong-Hongkong-Macau Greater Bay Area, SYSU, Zhuhai, 519082, China\\
}
\date{Accepted XXX. Received YYY; in original form ZZZ}

\pubyear{2015}

\begin{document}
\label{firstpage}
\pagerange{\pageref{firstpage}--\pageref{lastpage}}
\maketitle

\begin{abstract}

Cosmic void has been proven to be an effective cosmological probe of the large-scale structure (LSS). However, since voids are usually identified in spectroscopic galaxy surveys, they are generally limited to low number density and redshift. We propose to utilize the clustering of two-dimensional (2D) voids identified using Voronoi tessellation and watershed algorithm without any shape assumption to explore the LSS. We generate mock galaxy and void catalogs for the next-generation Stage IV photometric surveys in $z = 0.8-2.0$ from simulations, develop the 2D void identification method, and construct the theoretical model to fit the 2D watershed void and galaxy angular power spectra. We find that our method can accurately extract the cosmological information, and the constraint accuracies of some cosmological parameters from the 2D watershed void clustering are even comparable to the galaxy angular clustering case, which can be further improved by as large as $\sim30\%$ in the void and galaxy joint constraints. 
This indicates that the 2D void clustering is a good complement to galaxy angular clustering measurements, especially for the forthcoming Stage IV surveys that detect high-redshift universe.
\end{abstract}


\section {introduction} \label{sec:intro}

Cosmic voids are the regions with low densities and large volumes in the large-scale structure (LSS) of the Universe. It can be an effective cosmological probe for exploring the LSS \citep[e.g.][]{2011IJMPS...1...41V,2016IAUS..308..493V,2019BAAS...51c..40P,chan2021volume,2023JCAP...05..031S,mauland2023void}.  In particular, the three dimensional (3D) voids have been widely used in the studies of the LSS based on the statistical properties, such as void size function and number counts \citep{sheth2004hierarchy,jennings2013abundance,contarini2021cosmic,2022A&A...667A.162C,contarini2023cosmological,2023MNRAS.522..152P,2024JCAP...10..079V,2024MNRAS.532.1049S,vnc,2025arXiv250107817S}. 
However, since 3D voids are generally identified in spectroscopic surveys, they have some limitations or disadvantages in the LSS study, e.g. low number densities with poor statistics and locating at low redshifts with small redshift coverage.
On the other hand, if we can properly identify 2D voids in photometric surveys and correctly model them, the usability of voids as a cosmological probe can be greatly improved. Although some attempts have been made to use the 2D voids for exploring the LSS \citep[e.g.][]{2017MNRAS.465..746S,2018MNRAS.476.3195C,2021MNRAS.500..464V,2023A&A...670A..47B,2023JCAP...08..010V,2024A&A...689A.171C}, they are still preliminary which assume spherical void shapes and relatively simple modeling.

\begin{figure*}
\centering
\subfigure{
\includegraphics[width=1\columnwidth]{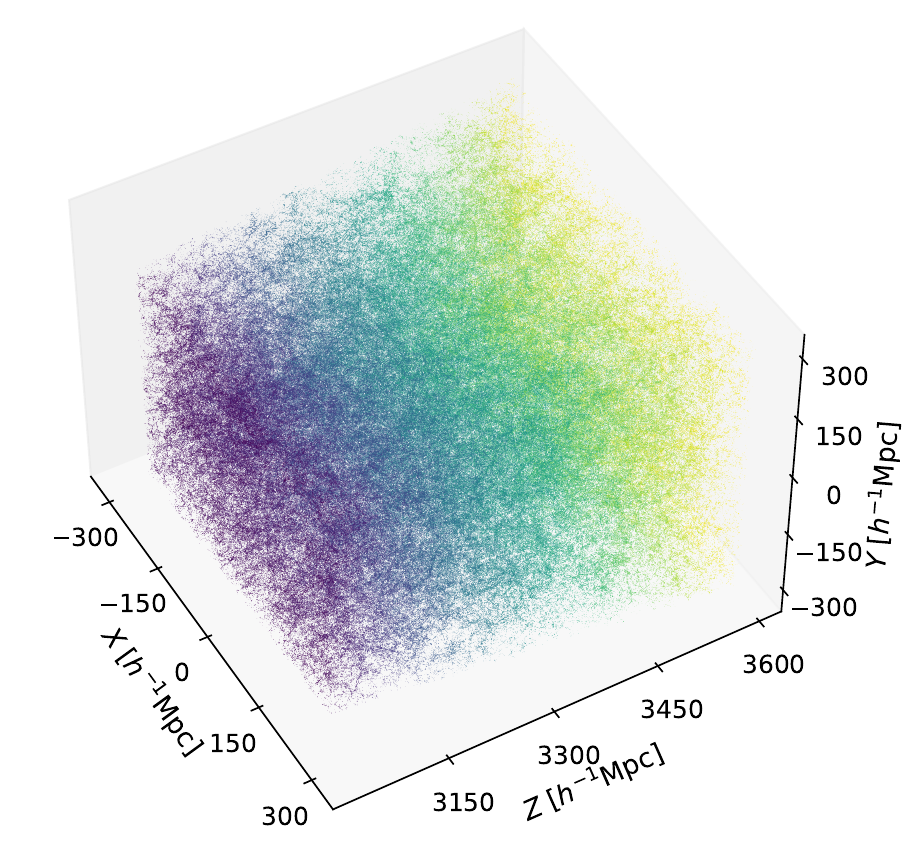}}
\hspace{2mm}
\subfigure{
\includegraphics[width=0.95\columnwidth]{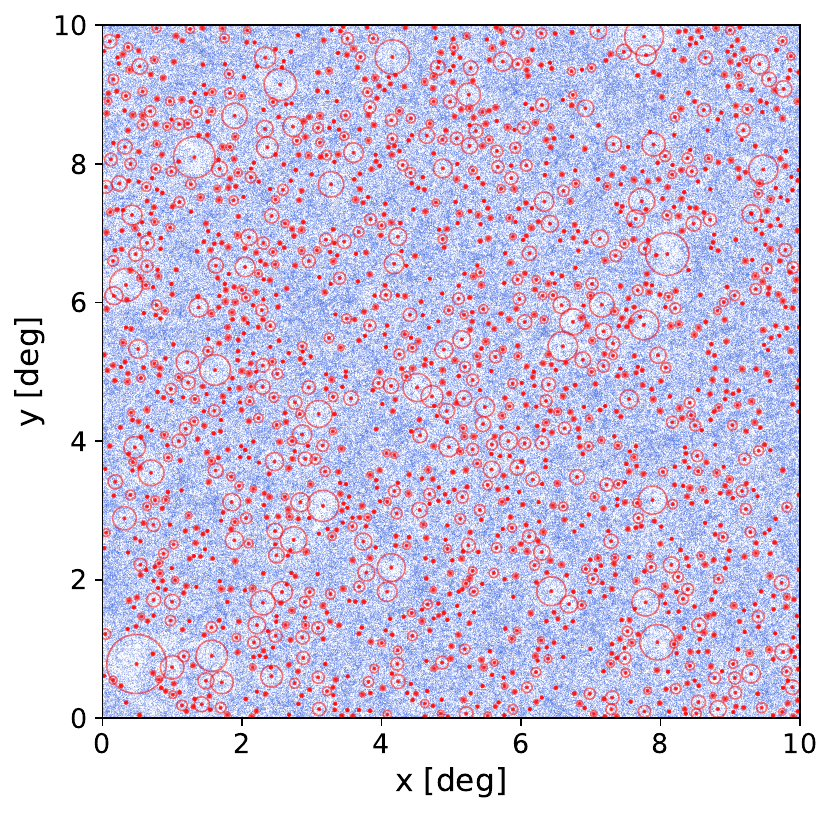}
}
\caption{As an example, we show the 3D spatial galaxy distribution (colored dots) in the left panel, and its 2D projection (blue dots) and the identified 2D voids (red dots) in the right panel for the photo-$z$ bin at $z=1.5-2.0$ in our simulations. The colored dots in the left panel, ranging from purple to yellow along the z axis, indicate the galaxies from low to high redshifts. The red circles and dots show the effective angular radius $\theta_{\rm v}$ and area-weighted centers $\mathbf{X}_{\rm v}$ of the 2D voids.}
\label{fig:catvis}
\end{figure*}

In this work, we propose a 2D void identification method based on the Voronoi tessellation \citep[e.g.][]{2009LNP...665..291V} and watershed algorithm \citep{watershed} without any shape assumption, and develop the corresponding theoretical model of the 2D watershed void clustering power spectrum using the halo model \citep{2014PhRvL.112d1304H}. The identification algorithm can find low surface density regions with natural and non-spherical shapes, and provide excellent 2D void clustering information for cosmological studies. This method can be applied to the next-generation Stage IV galaxy surveys, such as the Legacy Survey of Space and Time (LSST) \citep{2019ApJ...873..111I}, $\it Euclid$ \citep{2022A&A...662A.112E}, Roman space telescopes (RST) \citep{RST}, and
the China Space Station Telescope (CSST) \citep{zhan11,zhan2021csst,gong,2023MNRAS.519.1132M}. 

Taking the CSST photometric galaxy survey as an example, we generate the mock galaxy and void catalogs based on simulations and the CSST survey strategy. Then we derive the mock data of the void and galaxy auto and cross angular power spectra in four tomographic photometric redshift (photo-$z$) bins from $z=0.8$ to $2.0$. The Markov Chain Monte Carlo (MCMC) method is employed to constrain the cosmological and void parameters for exploring the feasibility and effectiveness of our method.

The paper is organized as follows: In Section \ref{sec:data}, we introduce the mock galaxy and 2D watershed void catalogs we use; In Section \ref{sec:aps}, we discuss the calculation of the theoretical model and generation of the mock data for the 2D void and galaxy clustering; In Section \ref{sec:mcmc}, we show the constraint results; We give our conclusion in Section \ref{sec:conclusion}.

\section{Mock Catalogs} \label{sec:data}
\subsection{Galaxy mock catalog} 
\label{sec:galcat}
We generate the mock galaxy catalogs from the dark matter-only Jiutian simulation, which contains $6144^3$ particles with a mass resolution of $m_{\rm p}$ = $3.72 \times 10^8$ $h^{-1}M_\odot$ and box size $1~ h^{-1}$Gpc. The fiducial cosmology we set are from $\it Planck$2018, i.e. $h = 0.6766$, $\Omega_{\text{m}} = 0.3111$, $\Omega_{\text{b}} = 0.0490$, $\Omega_\Lambda = 0.6899$, $\sigma_8 = 0.8102$ and $n_{\text{s}} = 0.9665$ \citep{2020A&A...641A...6P}.  
The friend-of-friend and subfind algorithm are used to identify dark matter halos and substructures \citep{2001NewA....6...79S,2005MNRAS.364.1105S}.
We use an updated version of the L-Galaxies semi-analytical model to place galaxies and construct a light cone covering 100 deg$^2$ sky area from $z=0$ to 3 \citep{2005MNRAS.364.1105S, 2006MNRAS.365...11C, 2007MNRAS.375....2D, 2011MNRAS.413..101G,henriques2015galaxy,2024MNRAS.529.4958P}. By tracking the merger tree of each galaxy, we can consider the evolution effect and naturally avoid galaxy repetition or omission at the boundary of slices.
Galaxies are selected based on the apparent magnitude limits of the CSST photometric survey, which can reach $i\sim26$ AB mag for 5$\sigma$ point source detection \citep{gong}. 

\begin{table}
    \center
    \caption{The galaxy and void surface number densities, i.e. $n_{\rm g}$ and $n_{\rm v}$ (in arcmin$^{-2}$), in the four photo-$z$ tomographic bins from $z=0.8$ to 2.0. The mean, minimum and maximum angular radii of voids $\theta_{\rm v}$ (in arcmin) with $R_{\rm v}>1\ h^{-1}\text{Mpc}$ and $D_{\rm cut}=0.2$ are also shown. }\label{tab:catalog}
    \begin{tabular}{ccccccc}
        \hline \hline
        $z_{\rm min}$ & $z_{\rm max}$ & $n_{\rm g}$ & $n_{\rm v}$ & $\theta_{\rm v}^{\rm mean}$ & $\theta_{\rm v}^{\text{min}}$ & $\theta_{\rm v}^{\text{max}}$\\
        \hline
        0.8 &1.0 & 3.59 & 0.0025 & 5.2 & 1.6 & 72.2\\
        1.0&1.2&2.86& 0.0022&3.7&1.5&31.0\\
        1.2&1.5&3.08& 0.0023&3.2&1.2&21.7\\  
        1.5&2.0&3.51& 0.0049&2.8&1.1&25.6\\ 
        \hline
	\end{tabular}
\end{table}

Since the LSS at low redshifts can be well measured by spectroscopic surveys, here we mainly focus on the high redshifts  at $z>0.8$. We also note that the galaxy density decreases quickly at $z>2$ in the CSST photometric survey \citep{gong}, which can dramatically suppress the surface density of 2D voids. Hence we only consider the redshift range $z= 0.8-2.0$ in our analysis, and split the galaxy sample into four photo-$z$ tomographic bins to extract more information and reduce 2D void overlapping effect. In Table~\ref{tab:catalog}, we show the galaxy surface densities $n_{\rm g}$ and redshift ranges for the four photo-$z$ bins we consider. The redshift range for each bin is determined to make $n_{\rm g}$ similar in each bin.
In Figure~\ref{fig:catvis}, as an example, we show the 3D spatial galaxy distribution and the 2D projection for the redshift bin $z=1.5 - 2.0$ in our simulations.

\begin{figure*}
    \centering

\subfigure{ \includegraphics[width=0.83\columnwidth]{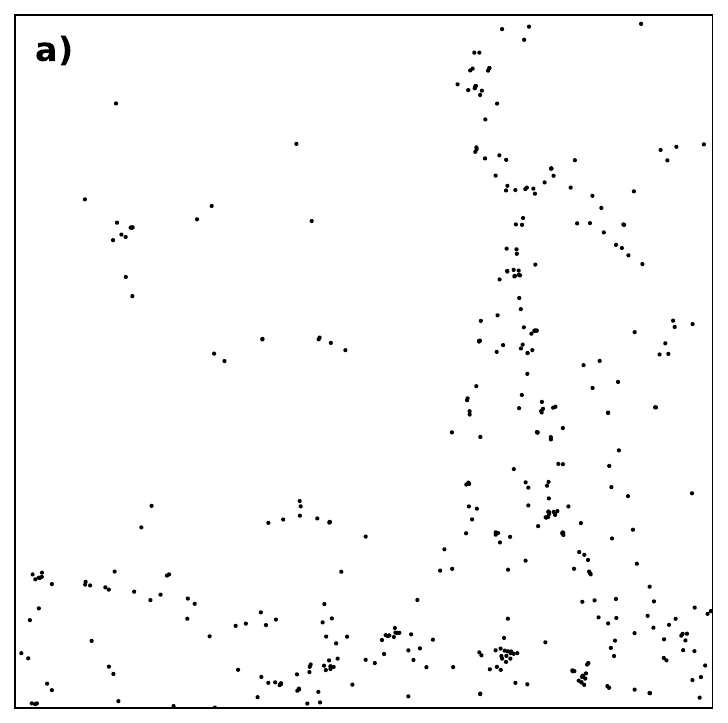}}
\hspace{3mm}
\subfigure{ \includegraphics[width=0.83\columnwidth]{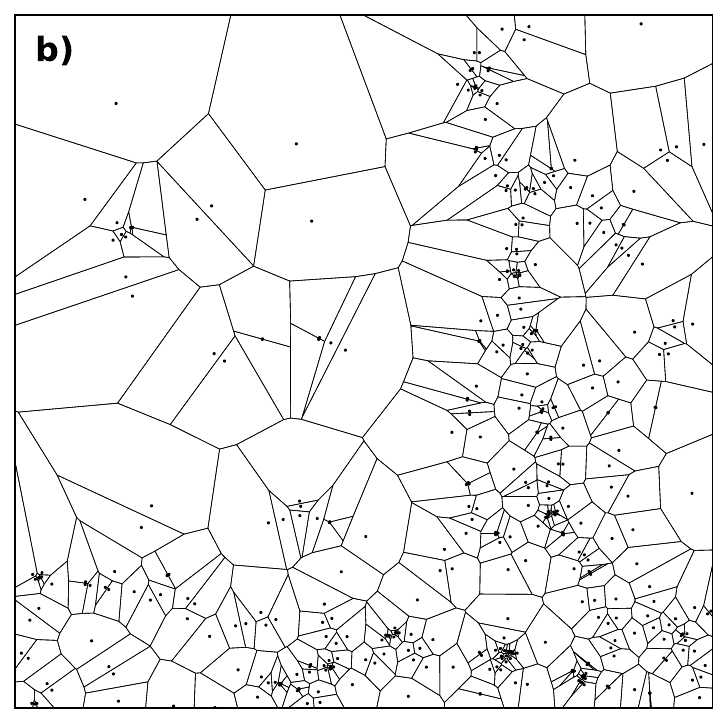}}

\subfigure{ \includegraphics[width=0.83\columnwidth]{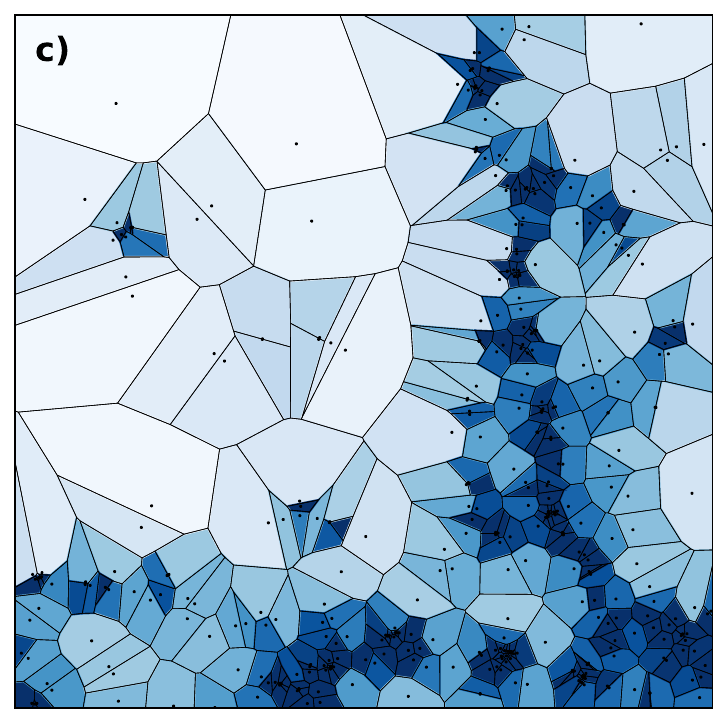}}
\hspace{3mm}
\subfigure{ \includegraphics[width=0.83\columnwidth]{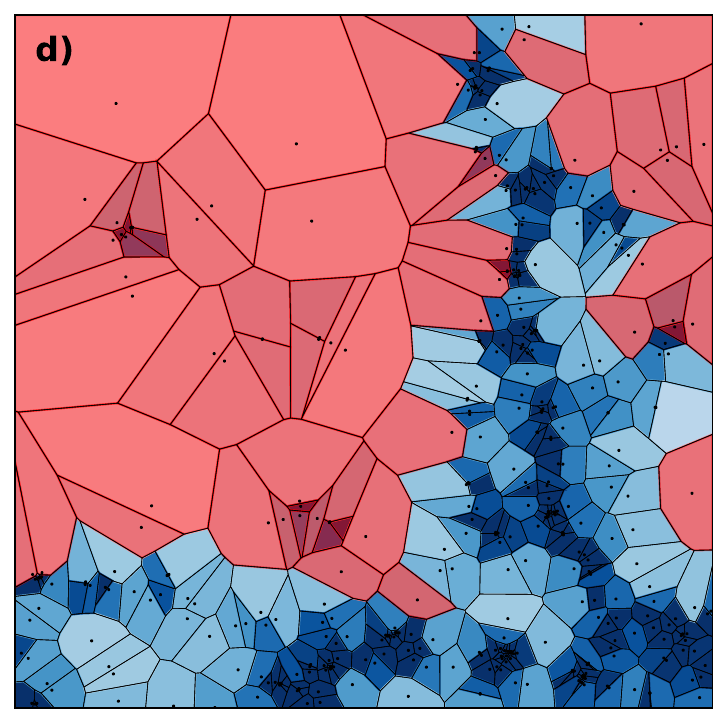}}
    \caption{The process of identifying 2D watershed voids from the galaxy catalog. \textbf{a):} Galaxies slice from a small part of the mock galaxy catalog. The black dots indicate the galaxy positions. \textbf{b):} The 2D Voronoi tessellation of galaxies in this slice. Each galaxy is assigned to a Voronoi cell. \textbf{c):} The cell of a galaxy is colored according to its area or density. \textbf{d):} 2D void candidates (red regions) are identified by the watershed algorithm without trimming.}\label{fig:vf}
\end{figure*}

\subsection{2D watershed void identification}
\label{sec:identification}

Here we provide a detailed description of the method we use to identify 2D watershed voids. We adopt the Voronoi tessellation and watershed algorithm to identify voids in 2D galaxy maps \citep{2020SciPy-NMeth,alphashape,2015opencv}, which have been widely used in 3D void identification. We show our detailed process for identifying the 2D void in Figure~\ref{fig:vf}. Our algorithm for finding the 2D void involves several steps.

First, we identify the cell of each tracer particle (i.e. galaxy, see Figure~\ref{fig:vf}a). The Voronoi tessellation allocates a cell to each galaxy based on the principle that each cell contains the region of space with a shorter distance to a galaxy than the distance to the galaxies in the other cells. Figure~\ref{fig:vf}b shows the result of applying Voronoi tessellation to a group of galaxies as an example.
Next, we estimate the density of a cell. The density is derived from the inverse of the cell area identified by Voronoi tessellation with $\rho_ {\rm cell}=1/S_ {\rm cell}$. We denote the densities using blue shades, with darker colors indicating higher densities, as shown in Figure~\ref{fig:vf}c. 

We then merge the cells into zones to form the 2D void candidates. This process is based on the watershed algorithm \citep{watershed,zobov}, which is inspired by the idea of water gradually flooding a landscape. This algorithm performs the topological delineation of regions around local minima in the density field by identifying boundaries at the ridges, which are defined by the connections between saddle points. To illustrate this process clearly, we can imagine a 2D density field as a water tank, where the water level represents the density of cells. As the tank fills with water, basins form around local minima, and their boundaries are defined by ridges where adjacent basins meet. 
We start with the low-density cells and merge the adjacent cells around them to form different zones. The maximum size that the zone can grow is considered as the boundary of a void candidate. This boundary is derived when the density of all the adjacent cells of the growing zone is greater than the average density $\bar{\rho}=\sum \rho_{\rm cell}/N_{\rm cell}$, where $N_{\rm cell}$ is the number of cells. These pool-like areas represent the 2D watershed void candidates. In Figure~\ref{fig:vf}d, we show the 2D void candidates generated by this process. 

\begin{figure*}
\subfigure{
\includegraphics[width=0.64\columnwidth]{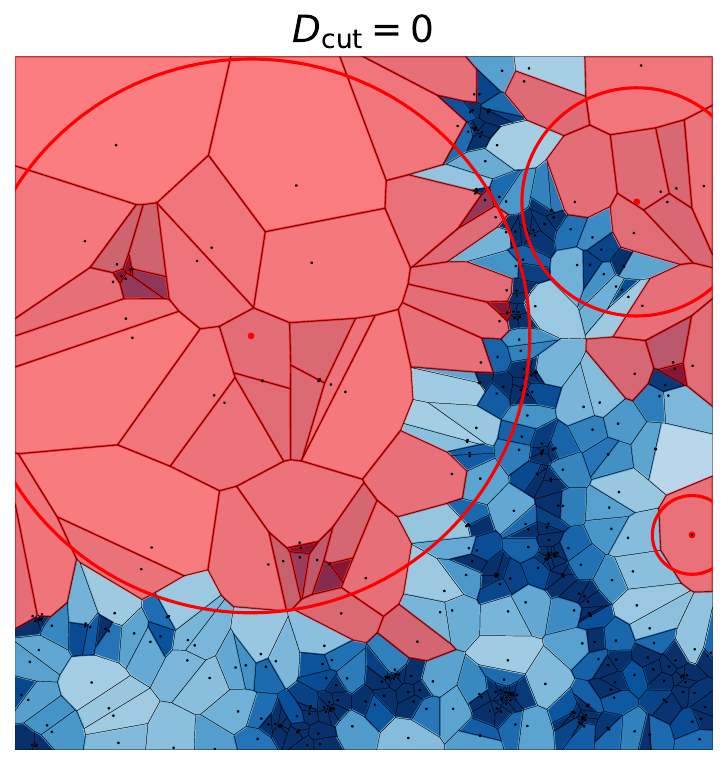}}
\hspace{2mm}
\subfigure{
\includegraphics[width=0.64\columnwidth]{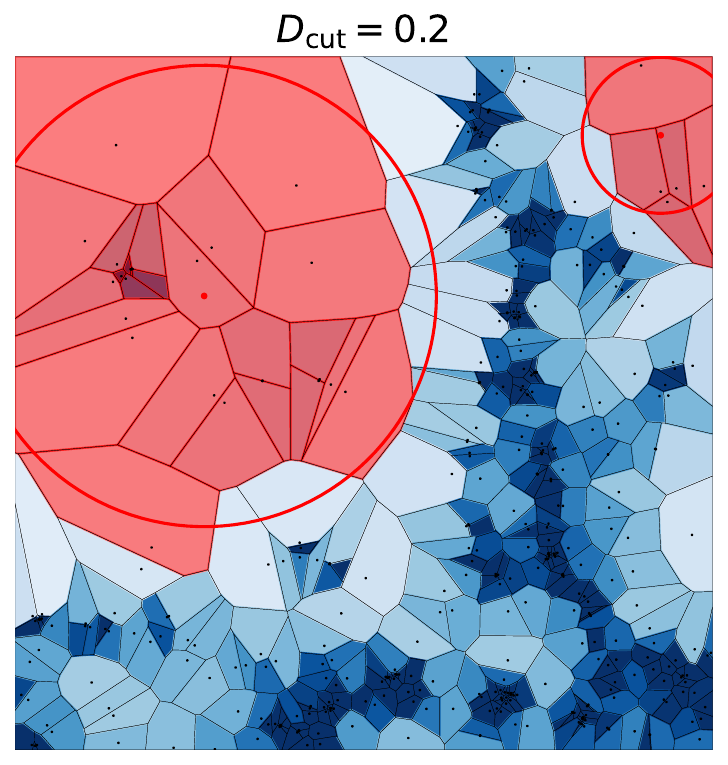}}
\hspace{2mm}
\subfigure{
\includegraphics[width=0.64\columnwidth]{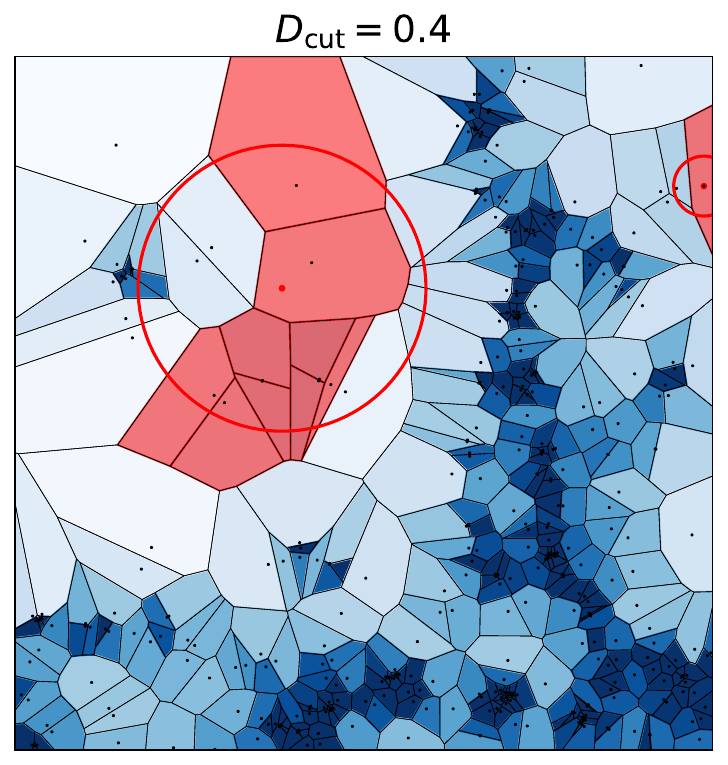}
}
\caption{The voids identified by choosing $D_{\rm cut}=0$ (left), 0.2 (middle), and 0.4 (right) in the trimming process. The density of each cell is shown in red or blue, and darker colors indicate higher densities. The 2D watershed voids are shown as red regions, and the red circles and dots denote the effective angular radius $\theta_{\rm v}$ and area-weighted centers $\mathbf{X}_{\rm v}$, respectively.}\label{fig:2dvoid}
\end{figure*}

Finally, we trim the boundaries of void candidates. This trimming step determines when to stop merging Voronoi cells during void identification. We find that applying a cut distance $D_{\rm cut}$ is more effective than simply adjusting the threshold for merging Voronoi cells by multiplying the average density $\bar{\rho}$ by a factor from the previous step. The value of $D_{\rm cut}$ determines how much of the boundary is trimmed. To estimate $D_{\rm cut}$, we divide the 2D galaxy map into a grid of pixels, with a resolution of 1000 by 1000 in our analysis, which is large enough for accurate estimation. The $D_{\rm cut}$ value is then derived by calculating the distance from the center of each pixel inside a void candidate to its nearest pixel on the boundary of the void candidate, and is normalized based on the maximum $D_{\rm cut}$ in the map, which takes the value from 0 to 1. The shape and size of a void candidate are trimmed by discarding the cells where $D_{\rm cut}$ is less than a certain value. This process is analogous to inserting a dam within the initial boundaries of pool-like void candidates, where $D_{\rm cut} = 0$ represents no trimming, i.e. retaining the original boundaries (see Figure~\ref{fig:vf}d). 
After these steps, we can obtain the 2D watershed voids from the 2D galaxy map. We discuss the $D_{\rm cut}$ selection and void mock catalog generation in the next subsection.


\subsection{Void mock catalog}
\label{sec:voidcat}
We identify mock 2D void catalogs from the galaxy mock catalog using the method described in Section~\ref{sec:identification}. To compute the void angular power spectrum, it is necessary to estimate their position and effective radius.

The void angular radius $\theta_{\rm v}$ is calculated from an effective circle with an area $S_{\rm v}$ equal to the total area of all cells within the void. Then the void area-weighted center $\mathbf{X}_{\text{v}}$ also can be estimated using the positions of the cells, and we have
\begin{equation}\label{eq:rv}
S_{\rm v}=\sum_i S^i_{\rm cell}=\pi \theta_{\rm v}^2, \  \ 
{\rm and}\ \mathbf{X}_{\text{v}} = \frac{1}{S_{\rm v}}\sum_i\mathbf{x}_i S^i_{\rm cell}.
\end{equation}
Here $S_{\text{cell}}^i$ represents the area of the cell $i$, and $\mathbf{x}_i$ is the coordinate of the galaxy within a cell in a given void.

To obtain a reliable void mock catalog for calculating the void angular power spectrum, as we mentioned above, our 2D void finder uses the cut distance $D_{\rm cut}$, which determines whether and to what degree the boundary of a void candidate is trimmed.
When $D_{\rm cut} = 0$, voids retain their original boundaries from the watershed process, which can result in oddly shaped voids, such as dumbbell-shaped or hook-like structures. These shapes can cause significant deviations in the calculation of void positions. On the other hand, if $D_{\rm cut}$ is too large, the number of voids decreases significantly, and the effective angular radii of the remaining voids become much smaller. In both cases, the void catalog does not accurately represent the true void distribution, thus affecting the calculation of the void angular power spectrum.

To evaluate the impact of different $D_{\rm cut}$ values, we calculate the fraction of Voronoi cells within each void that fall inside its effective angular radius, i.e. the number of cells in a void that fall inside the void effective radius divided by the total number of cells belonging to that void. Hence, a fraction closer to 1 indicates a more accurate identification. We find that, when $D_{\rm cut} = 0.2$,  the number of voids with this fraction equal to 1 is the largest, compared to other $D_{\rm cut}$ values. This indicates that $D_{\rm cut} = 0.2$ can provide the most accurate identification of 2D voids.

In Figure~\ref{fig:2dvoid}, we show the effects of identifying 2D watershed voids by selecting different $D_{\rm cut}$ values. We can visually find that the identified 2D void (red regions) and the effective angular radius $\theta_{\rm v}$ (red circle) are well matched when $D_{\rm cut}$ = 0.2, which shows that $\theta_{\rm v}$ and $\mathbf{X}_{\rm v}$ are calculated more accurately in this case. So we choose $D_{\rm cut}$ = 0.2 to trimming 2D void candidates in our analysis. Besides, to reduce the non-linear evolution effect and the incompleteness of the small-size voids, we remove the voids with the effective radius $R_{\rm v}=\theta_{\rm v} D_{\rm A}<1\ h^{-1}{\rm Mpc}$, where $D_{\rm A}$ is the comoving angular diameter distance.


In Table~\ref{tab:catalog}, we show the 2D void surface number densities, and the average, minimum and maximum angular radii of 2D voids with $D_{\rm cut}=0.2$ and $R_{\rm v}>1\ h^{-1}\text{Mpc}$  in the four photo-$z$ tomographic bins.
We find that the mean void angular radius $\theta^{\rm mean}_{\rm v}$ has a trend to become smaller as redshift increases, while the mean void radius $R^{\rm mean}_{\rm v}$ is similar ($\sim 3\ h^{-1}\text{Mpc}$) at $z<2$. In Figure~\ref{fig:catvis}, we show the area-weighted centers and effective angular radii of the identified 2D voids with $D_{\rm cut}$ = 0.2 and $R_{\rm v}>1\ h^{-1}\text{Mpc}$ at $z=1.5 - 2.0$.

\begin{figure*}
    \centering
	\subfigure{\includegraphics[width=1\columnwidth]{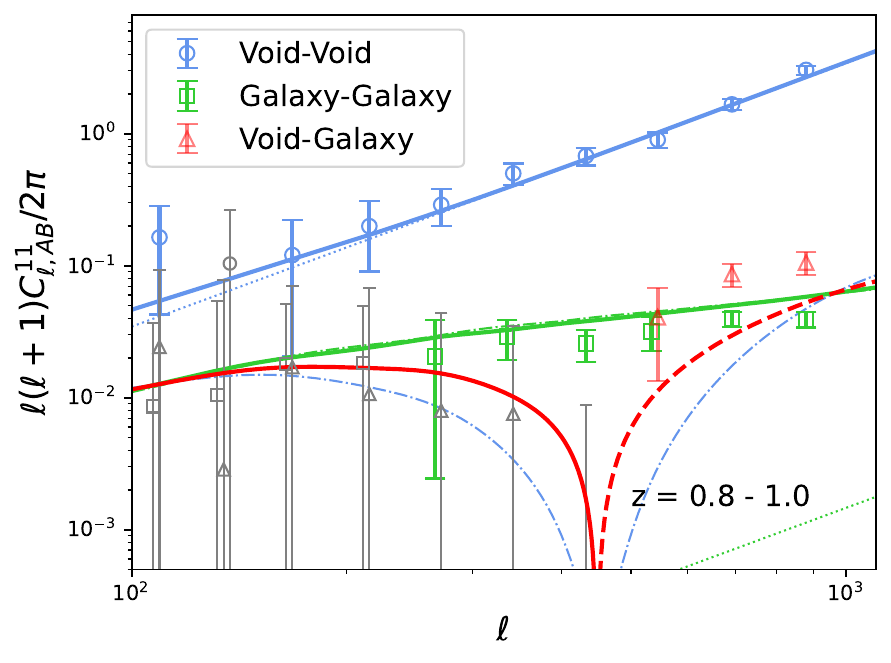}}
    \hspace{5mm}
    \subfigure{\includegraphics[width=1\columnwidth]{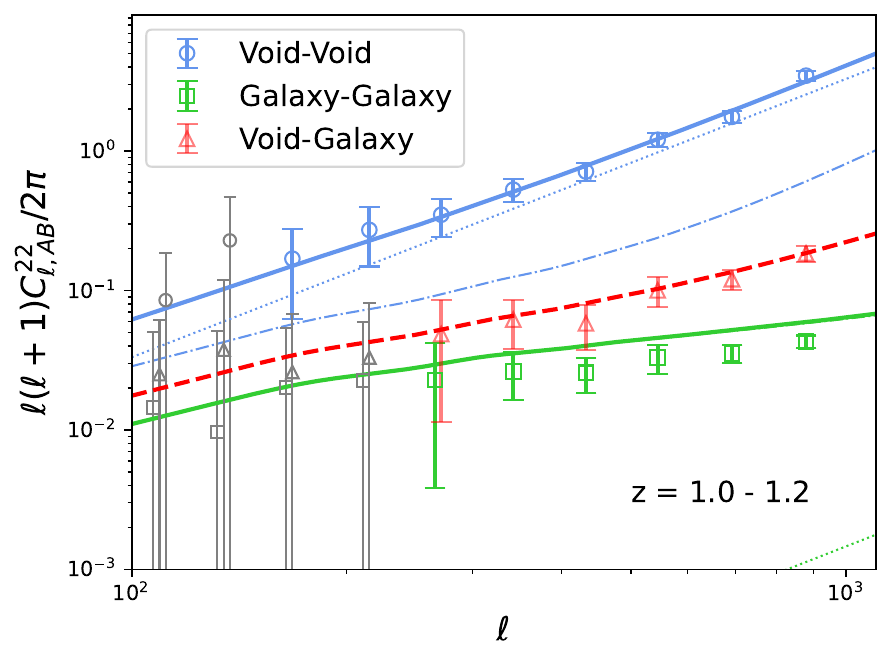}}
    \subfigure{\includegraphics[width=1\columnwidth]{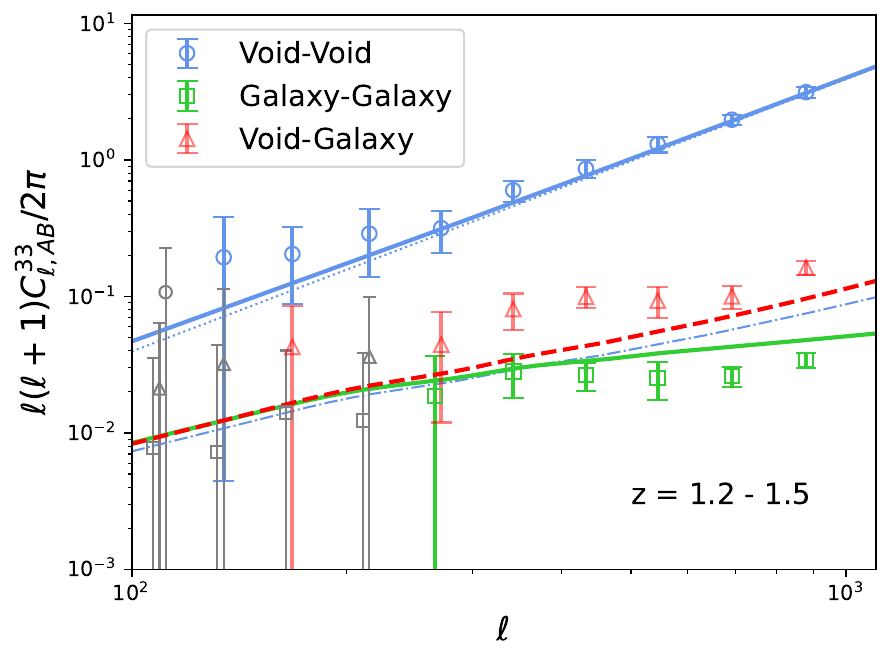}}
    \hspace{5mm}
    \subfigure{\includegraphics[width=1\columnwidth]{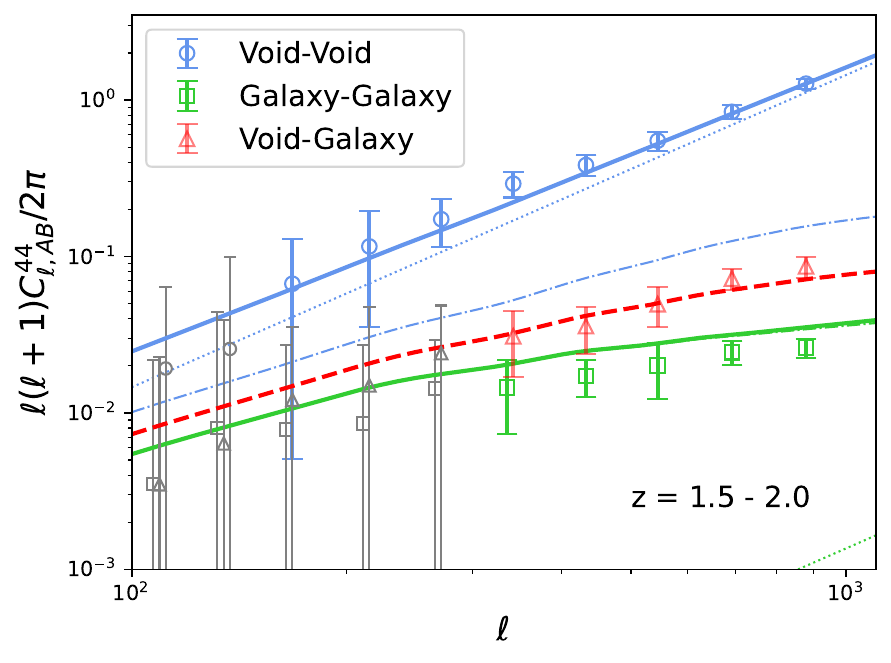}}
    \caption{The angular power spectra $C_{\rm vv}^{ij}$ (blue), $C_{\rm gg}^{ij}$ (green) and $C_{\rm vg}^{ij}$ (red) with $i=j$ in the four rehshift tomographic bins from the 100 deg$^2$ simulation. The gray data points with SNR $<$ 1 are excluded from the constraint process. The curves are the best-fits of the theoretical calculation. The red dashed lines and data points indicate that the values are negative. The solid (or dashed), dash-dotted and dotted curves are the total, clustering and noise terms of the power spectra, respectively.}\label{fig:cl}
\end{figure*} 

\section{Angular Power Spectrum}\label{sec:aps}
We model the galaxy and void auto and cross angular power spectrum based on the halo model \citep{2014PhRvL.112d1304H}. Assuming Limber approximation \citep{1954ApJ...119..655L}, the galaxy and void angular power spectra can be written as
\begin{equation}\label{eq:cl}
    C_{\rm AB}^{ij}(\ell)=\frac{1}{c}\int\frac{H(z)}{D_A^2(z)}W^i_{\rm A}(z)W^j_{\rm B}(z)P_{\rm mm}\left[\frac{\ell+1/2}{D_{\rm A}(z)},z\right]dz,
\end{equation}
where A and B denote two different tracers, $c$ is the speed of light, and $P_{\rm mm}$ is the matter power spectrum, where the wavenumber $k$ is converted by the multipole $\ell$ and comoving angular diameter distance $D_{\rm A}$. In this work, we use \texttt{CAMB} \citep{camb} to calculate $P_{\rm mm}$ and $D_{\rm A}$. 

The galaxy weighting function is given by $W_{\rm g}^i(z)=b_{\rm g}^in_{\rm g}^i(z)$,
where $b^i_{\rm g}$ is the galaxy bias, and $n^i_{\rm g}(z)$ is the normalized galaxy redshift distribution in the $i$th bin. For the void weighting function, we include the void density profile in the model and find that
\begin{equation}
    W_{\rm v}^i(z)=b_{\rm v}^in_{\rm v}^i(z)u^i_{\rm v}\left[k=\frac{\ell+1/2}{D_{\rm A}(z)},z\right].
\end{equation}
Here $b^i_{\rm v}$ is the void bias, and $n^i_{\rm v}(z)$ is the normalized void redshift distribution in the $i$th bin. $u_{\rm v }(k)$ is the normalized void density profile in Fourier space \citep{2014PhRvL.112d1304H}
\begin{equation}\label{eq:uvk}
u_{\rm v}(k)=\frac{\bar{\rho}}{\delta m}\int_{0}^{\infty}{u_{\rm v}(r)\frac{\sin (kr)}{kr}4\pi r^2}dr.
\end{equation}
Here $\delta m$ is the void uncompensated mass, which is given by $\delta m=\bar{\rho}\int_{0}^{\infty}{u_{\rm v}(r)4\pi r^2}dr$, and $u_{\rm v}(r)$ is the averaged spherically deviation between the void mass density and the mean matter density of the entire universe. 
It can be derived from the form in the real space \citep{2014PhRvL.112y1302H}
\begin{equation}\label{eq:vdp}
u_{\rm v}(r)=\frac {\rho_{\rm v}(r)} {\bar{\rho}}-1=\delta_{\rm cen} \frac {1-(r /R_{\rm s})^\alpha} {1+(r /R_{\rm v})^\beta}.
\end{equation}
Here $R_{\rm s}\equiv \gamma R_{\rm v}$ is the scale radius when the void density $\rho_{\rm v}$ = $\bar{\rho}$, where $\gamma$ is the ratio factor and $\bar{\rho}$ is the mean matter density. $\delta_{\rm cen}=-1$ is the central density contrast, which can be cancelled out in our analysis (see Eq.~(\ref{eq:uvk})). $\alpha$ and $\beta$ denote the inner and outer slopes of the compensation wall around a void, and we set them as free parameters in our fitting process. 
Here we propose to simplify the calculation by setting $R_{\rm v}=R_{\rm v}^{\rm mean}$ in Eq.~(\ref{eq:vdp}). This means that $u_{\rm v}(r)$ and $u_{\rm v}(k)$, as well as $R_{\rm s}$, $\alpha$, $\beta$ and $\gamma$, are the mean values of all voids selected in a redshift bin. 

When we calculate the angular power spectra of galaxy, void and galaxy-void, statistical and systematical uncertainties are also considered by adding the noise term $\widetilde{C}_{\rm AB}=C_{\rm AB}+N_{\rm AB}$, where $N_{\rm AB}$ contains  the shot noise and systematics.
Besides the three parameters from the void density profile, we also set the galaxy bias $b_{\rm g}$, void bias $b_{\rm v}$, and noise terms as free parameters when constraining the cosmological parameters. 

We use \texttt{powerbox} \citep{2018JOSS....3..850M} to derive the data of the void and galaxy angular power spectra from the mock catalogs, and estimate errors using the jackknife method. We only use the data points with the signal-to-noise ratio (SNR) $>$ 1 to obtain sufficient statistical significance in the fitting process. In Figure~\ref{fig:cl}, we show the mock data of the galaxy, void and void-galaxy angular power spectra at four redshift bins in the 100 deg$^2$ survey area from the simulation. Note that we only consider the void and galaxy angular power spectra in the same redshift bin for simplicity, since the cross power spectra between different photo-$z$ bins are relatively small with large errors, especially for voids. In the full Stage IV surveys with more than ten thousand survey area, these cross power spectra need to be considered. 

\begin{figure}
	\includegraphics[width=\columnwidth]{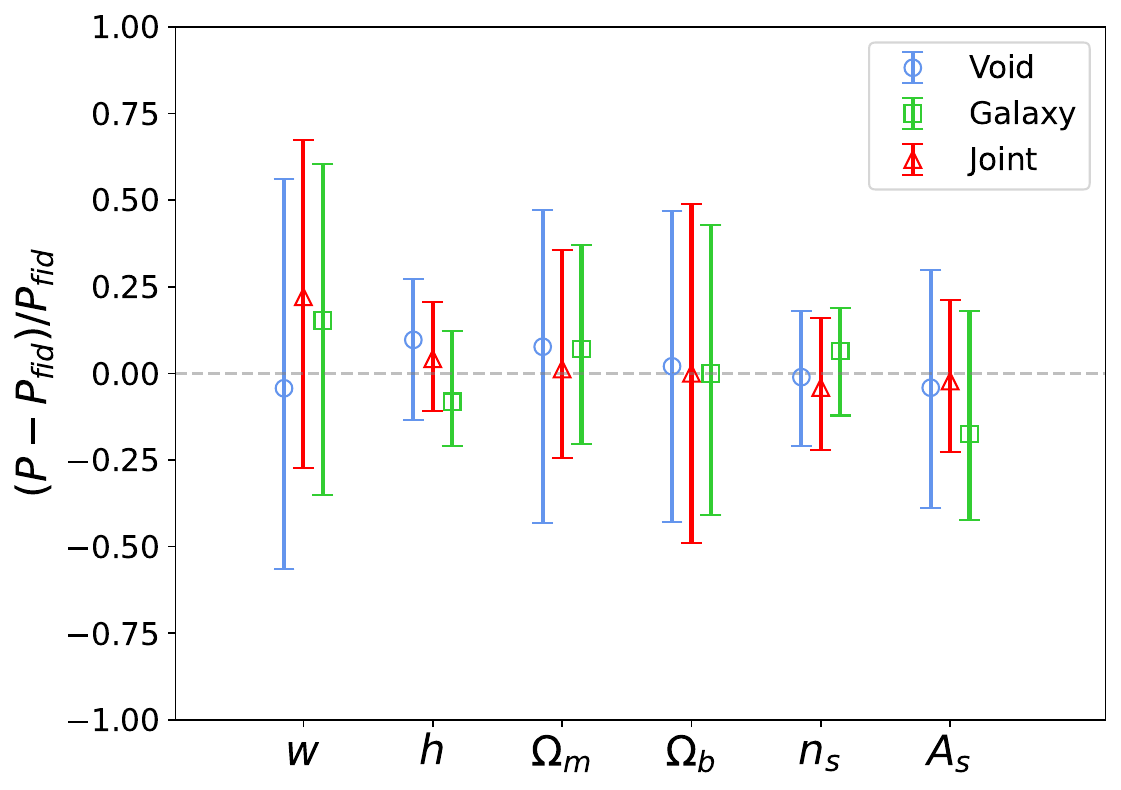}
    \caption{The residuals between the best-fit and fiducial values of the cosmological parameters from the simulation covering 100 deg$^2$ . The error bars denote the 1$\sigma$ constraint results.}
    \label{fig:mcmccosmic}
\end{figure}

\section{Parameter Constraint}\label{sec:mcmc}
We constrain the model parameters by $\chi^2$, which takes the form as
\begin{equation}
    \chi_{\rm AB}^2 =       \sum\left[\frac{C_{\rm AB}^{\rm data}(\ell)-C_{\rm AB}^{\rm th}(\ell)}{\sigma^{\rm data}}\right]^2,
\end{equation}
where $\sigma^{\rm data}$ is the error of the mock data, $C_{\rm AB}^{\rm th}(\ell)$ and $C_{\rm AB}^{\rm data}(\ell)$ are the theoretical and the mock data angular power spectrum, respectively. 
We constrain the model parameters using $\chi^2$ method, and the likelihood function can be derived by $\mathcal{L}$ $\propto$ exp($-\chi^2$/2). The total $\chi^2$ for the joint constraints with all power spectra in the four redshift bins can be calculated by $\chi^2_{\rm tot} = \chi^2_{\rm vv} + \chi^2_{\rm gg} + \chi^2_{\rm vg}$, where $\chi^2_{\rm vv}, \chi^2_{\rm gg}$ and $\chi^2_{\rm vg}$ are the chi-squares for the void, galaxy and void-galaxy angular power spectrum. The MCMC is employed to constrain the free parameters, which is performed by \texttt{emcee} \citep{emcee} code. We choose 112 walkers and obtain 30000 steps for each chain. The first 10 percent of steps are discarded as the burn-in process. 

In the parameter constraint process, we assume flat priors, and have included 6 cosmological parameters, 
i.e. $w\in(-1.8,-0.2)$, $h\in(0.5,0.9)$, $\Omega_{\rm m}\in(0.1,0.5)$, $\Omega_{\rm b}\in(0.02,0.08)$, $A_{\rm s}/10^{-9}\in(1,3)$, $n_{\rm s}\in(0.7, 1.2)$, 
and 3 void parameters $\alpha^i\in(0,10)$, $\beta^i\in(0,20)$, $\gamma^i\in(0,2)$ in each redshift bin. The galaxy and void biases $b_{\rm g}^i \in (0,5)$ and $b_{\rm v}^i \in (-20,20)$, and noise terms ${\rm log}_{10}(N^i_{\rm g}) \in (-20,0)$, ${\rm log}_{10}(N^i_{\rm v}) \in (-20,0)$ and $N^i_{\rm vg}/10^{-8} \in (-1000,1000)$ in each redshift bin are also considered.  Note that $N_{\rm vg}$ can have both positive and negative values since it contains the statistical uncertainty from measuring the void density profile \citep{2014PhRvL.112d1304H}. We also consider the 1$\sigma$ constraint results from the galaxy angular power spectra only, and set tighter prior ranges for $b_{\rm g}^i$  
to obtain better constraint results in the joint fitting case.

The best-fit curves of the theoretical model in the fourth redshift bin are shown in Figure~\ref{fig:cl}. 
We can find that the theoretical curves can fit the 2D watershed void, galaxy and void-galaxy power spectra well.
In Figure~\ref{fig:mcmccosmic}, we show the residuals between the best-fit and fiducial values. We can see that the fitting results of the cosmological parameters are consistent with the fiducial values within 1$\sigma$ confidence level (CL). 
These indicate that our 2D void identification method can effectively produce the 2D void catalog, and our theoretical model can accurately extract the cosmological information from the 2D watershed void clustering.

We also notice that although the noise term is large in the 2D void power spectrum mainly due to the relatively low surface number density, the constraint powers for some cosmological parameters, e.g. $h$ and $A_{\rm s}$, are comparable to the galaxy angular power spectrum, and $w$ and $n_{\rm s}$ are only slightly worse. Besides, the constraint accuracy can be further improved as large as $\sim30\%$ in the joint constraint case, compared to the galaxy only case. 
This indicates that the 2D watershed void clustering can complement the galaxy clustering measurement for extracting cosmological information as an effective cosmological probe.

\begin{figure}
	\includegraphics[width=\columnwidth]{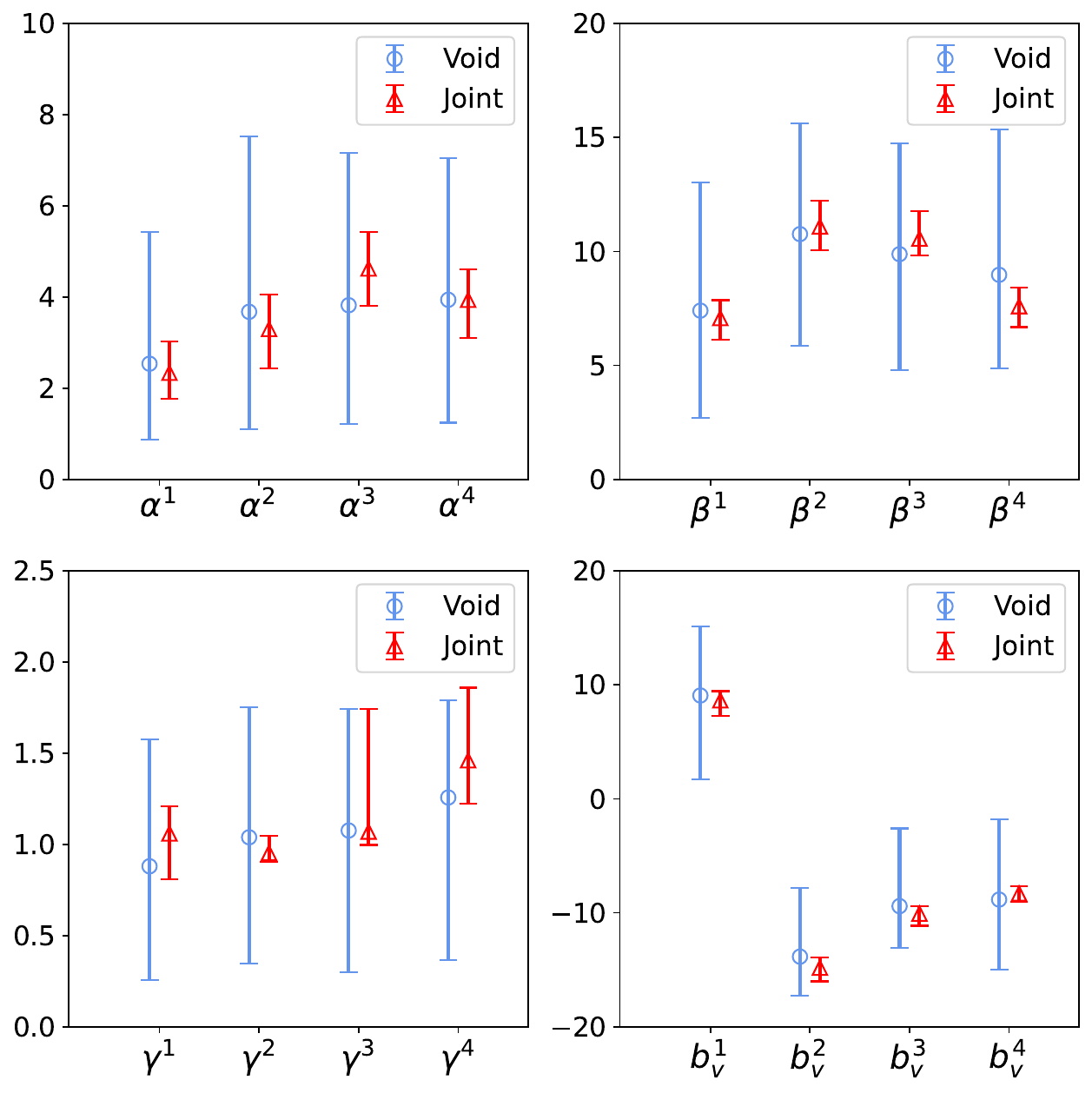}
    \caption{The best-fit and 1$\sigma$ constraint results of $\alpha^i$, $\beta^i$, $\gamma^i$ and $b_{\rm v}^i$ from the void only (blue) and joint constraints (red).}
    \label{fig:mcmcvp}
\end{figure}

In Figure \ref{fig:mcmcvp}, we show the best-fits and 1$\sigma$ CL of $\alpha$, $\beta$, $\gamma$ and $b_{\rm v}$ from the void power spectrum only  and joint constraints. We can find that the constraint results are consistent within 1$\sigma$ for these two cases, and the results from the joint constraints improve significantly by considering the galaxy and void-galaxy power spectra. We also notice that the best-fit values of  $\alpha^i$, $\beta^i$ and $\gamma^i$ have no obvious evolution trend with redshift, and basically fluctuate around certain values, i.e. $\alpha\sim3.5$, $\beta\sim9$ and $\gamma\sim1$.
For the void bias $b_{\rm v}$, the best-fit value is positive in the first redshift bin with $z<1$, and negative at $z>1$. And we note that the probability distributions of $b_{\rm v}$ can cover both positive and negative values when only considering the 2D void power spectrum. And the joint constraints can further improve the accuracy by about several times or even one order of magnitude on $b_{\rm v}^i$ and as large as 80\% on $\alpha^i$, $\beta^i$ and $\gamma^i$. These results are also consistent with the previous 3D void studies \citep[e.g.][]{2019MNRAS.490.3573F,2024ApJ...976..244S}. 

In addition, the other systematical parameters, such as $b_{\rm g}$, $N_{\rm g}$, $N_{\rm v}$ and $N_{\rm vg}$, are also well constrained in our analysis. We obtain $b_{\rm g}^1=1.560_{-0.181}^{+0.227}$, $b_{\rm g}^2=1.724_{-0.192}^{+0.257}$, $b_{\rm g}^3=2.095_{-0.265}^{+0.257}$, $b_{\rm g}^4=2.480_{-0.217}^{+0.333}$ in the joint fitting process. For the noise term of the void and galaxy angular power spectrum, we have similar constraint results as ${\rm log}_{10}(N^i_{\rm v}) \simeq -5$ and ${\rm log}_{10}(N^i_{\rm g}) \simeq -8$ at $i$th redshift bins, which give the constraint accuracies as ${\rm log}_{10}(N^i_{\rm v}) \sim 1\%$ and ${\rm log}_{10}(N^i_{\rm g}) \sim 10\%$ in the joint fitting process. And we have the constraint results of the noise from void-galaxy angular power spectrum as $N_{\rm vg}^1/10^{-8}=-0.594_{-1.188}^{+1.380}$, $N_{\rm vg}^2/10^{-8}=1.187_{-0.886}^{+1.078}$, $N_{\rm vg}^3/10^{-8}=-29.733_{-0.999}^{+0.982}$, $N_{\rm vg}^4/10^{-8}=0.546_{-1.475}^{+1.070}$ in the joint fitting process. We find that the joint constraints can provide a $\sim$50\% improvement on $b^i_{\rm g}$ and $\sim$70\% on $N^i_{\rm g}$ at $i$th redshift bins. 
Since the noise term $N_{\rm v}$ is dominated in measuring the void angular power spectrum, we can obtain tight constraint on $N_{\rm v}$ from both void angular power spectrum only and joint constraint, which are consistent within 1$\sigma$.

\section{conclusion}
\label{sec:conclusion}
We propose to use the clustering of 2D watershed voids with natural and non-spherical shapes to explore the LSS. We develop a 2D void identification method based on the Voronoi tessellation and watershed algorithm, and build a theoretical model to extract the cosmological and void information. By generating the galaxy and void mock catalog using Jiutian simulations for the CSST photometric survey covering 100 deg$^2$ from $z=0.8-2.0$, we study the feasibility of this method.

We find that the watershed void and void-galaxy angular power spectrum can accurately derive the cosmological information, and the best-fit values of the cosmological parameters are consistent with the fiducial values within 1$\sigma$ CL. The constraint strength of 2D void clustering for some cosmological parameters is comparable to the galaxy angular clustering, and the constraint accuracy can be improved as large as $\sim30\%$ in the joint fitting case. These indicate that the 2D watershed void clustering can be an effective cosmological probe and a good complement to the galaxy photometric survey, especially for the upcoming Stage IV surveys to probe the high-$z$ universe.

\section*{Acknowledgements}

YS and YG acknowledge the support from National Key R\&D Program of China grant Nos. 2022YFF0503404, 2020SKA0110402, and the CAS Project for Young Scientists in Basic Research (No. YSBR-092). KCC acknowledges the support the National Science Foundation of China under the grant number 12273121. XLC acknowledges the support of the National Natural Science Foundation of China through Grant Nos. 12361141814, and the Chinese Academy of Science grants ZDKYYQ20200008. QG acknowledges the support from the National Natural Science Foundation of China (NSFC No.12033008). This work is also supported by science research grants from the China Manned Space Project with Grant Nos. CMS-CSST-2021-A01 and CMS-CSST-2021-B01.

\section*{Data Availability}

 The data that support the findings of this study are available from the corresponding author upon reasonable request.



\bibliographystyle{mnras}
\bibliography{2dref} 




\appendix


\bsp	
\label{lastpage}
\end{document}